\documentclass[sigconf]{acmart}

\usepackage{color}
\usepackage{soul}

\setcopyright{none}
\renewcommand\footnotetextcopyrightpermission[1]{}
\begin{document}

\title{SCHeMa: Scheduling Scientific Containers on a Cluster of Heterogeneous Machines}


\author{Thanasis Vergoulis}
\email{vergoulis@athenarc.gr}
\orcid{0000-0003-0555-4128}
\affiliation{%
  \institution{IMSI, ``Athena'' RC}
}

\author{Konstantinos Zagganas}
\email{kzagganas@uop.gr}
\affiliation{%
  \institution{Univ. of the Peloponnese \& IMSI, ``Athena'' RC}
}

\author{Loukas Kavouras}
\email{kavouras@athenarc.gr}
\affiliation{%
  \institution{IMSI, ``Athena'' RC}
}

\author{Martin Reczko}
\email{reczko@fleming.gr}
\affiliation{%
  \institution{B.S.R.C. ``Al. Fleming''}
}

\author{Stelios Sartzetakis}
\email{stelios@athenarc.gr}
\affiliation{%
  \institution{IMSI, ``Athena'' RC \& GRNET}
}

\author{Theodore Dalamagas}
\email{dalamag@athenarc.gr}
\affiliation{%
 \institution{IMSI, ``Athena'' RC}
}

\renewcommand{\shortauthors}{Vergoulis, et al.}

\begin{abstract} 
In the era of data-driven science, conducting computational experiments that involve analysing large datasets using heterogeneous computational clusters, is part of the everyday routine for many scientists. Moreover, to ensure the credibility of their results, it is very important for these analyses to be easily reproducible by other researchers. Although various technologies, that could facilitate the work of scientists in this direction, have been introduced in the recent years, there is still a lack of open source platforms that combine them to this end. In this work, we describe and demonstrate SCHeMa, an open-source platform that facilitates the execution and reproducibility of computational analysis on heterogeneous clusters, leveraging containerization, experiment packaging, workflow management, and machine learning technologies. 
\end{abstract}

\begin{CCSXML}
<ccs2012>
<concept>
<concept_id>10002951.10003227.10010926</concept_id>
<concept_desc>Information systems~Computing platforms</concept_desc>
<concept_significance>300</concept_significance>
</concept>
</ccs2012>
\end{CCSXML}

\ccsdesc[300]{Information systems~Computing platforms}

\keywords{containerization, data-driven science, reproducibility}

\maketitle

\section{Introduction}
\label{sec:intro}

The computational analysis of large datasets has been established as an important part of the daily routine of scientists in many disciplines, shaping the field of \emph{data-driven science}. Due to the large size of the datasets, such computations are assigned to the nodes of computational clusters owned by the academic or research institution to which the scientists belong. It is very common that such computational clusters are heterogeneous, consisting of machines of very diverse specifications (CPUs, memory, disk, etc.) or capabilities (e.g., support for FPGAs and other accelerators). This \emph{heterogeneity} is due to the fact that these infrastructures (a)~have to serve a variety of analysis tasks, each having its own special needs (e.g., to exploit accelerators), and (b)~are usually built incrementally, with equipment units being procured at different (and maybe significantly distant) time periods, based on the availability of funds.

Similarly to any other scientific experiment, replicating and reproducing the results of a computational analysis is an important guarantee for its credibility. This is especially important nowadays, since there is an increasing concern in the research and academic community about the existence of a large number of scientific works that cannot be reproduced~\cite{baker2016}. Although it may be an exaggeration that we experience an ongoing \emph{reproducibility crisis}, it is unarguable that this is an important phenomenon that needs to be addressed~\cite{Fanelli2628}. This is why facilitating reproducibility has become an important topic of many research and academic disciplines.

In the context of data-driven science, facilitating reproducibility can be translated into making the datasets, the code, and the configurations used for the analysis openly available. Motivated by this need, approaches to pack up scientific computational experiments (e.g., RO-crate~\cite{ro-crate}) have been introduced. Although such packages are really useful, their true potential is not easily unleashed due to the fact that computer environments (e.g., software libraries, packages, programming languages) are complex and rapidly evolving, making the reproducibility and extension of computational analyses challenging and tedious~\cite{Boettiger14}. For instance, although the code of a computational experiment may be openly available (e.g., on GitHub or other similar repositories), in many cases its installation may require searching for old or, even, deprecated versions of third-party software, resolving conflicts between different versions of particular dependencies required by different software units, adapting the code to work on a different operating system, and so on. Given the fact that, usually, scientific software lacks comprehensive documentation, tasks like the previous may require significant technical skills that most scientists do not possess.  

The missing piece in this puzzle is the use of \emph{containerization technologies} (e.g., Docker, Singularity), along with experiment packaging, which has the potential to alleviate issues like these~\cite{Boettiger14,chamberlain2014,jimenez2015}. Such technologies allow the code of a complex software unit to be packed up with its dependencies so it can be easily and reliably executed in a variety of computing environments (laptops, PCs, Cloud or HPC nodes, etc). This packaged software units are known as \emph{container images}, and can further facilitate the reproducibility of computational analysis tasks since they are already configured and ready-to-use without requiring advanced technical skills for their installation. In addition, in the last years, a large number of scientific containers have been released in public repositories (e.g., in the time of writing, Biocontainers~\cite{da2017biocontainers} currently contain more than $2076$ containers).

It is evident, from the previous discussion, that various technologies, that could facilitate the work of scientists in the direction of data-driven science, have been introduced in the recent years. Consequently, various platforms that attempt to combine these technologies to provide useful services for the research community exist. For example, two relevant platforms are EOSC Life's WorkflowHub\footnote{WorkflowHub: \url{https://workflowhub.eu/}} and Galaxy Europe\footnote{Galaxy Europe: \url{https://usegalaxy.eu/}}~\cite{galaxy}. The former is an under-development, federated repository of workflows that is based on the SEEK platform~\cite{SEEK2015}, however it does not support workflow execution by itself. The latter is a feature-rich platform that supports job scheduling on heterogeneous clusters.

In this work, we introduce SCHeMa (\underline{S}cheduler for scientific \underline{C}ontainers on clusters of \underline{He}terogeneous \underline{Ma}chines) a new open source\footnote{SCHeMa's code repository: \url{https://github.com/athenarc/schema} (GNU/GPL license)} platform that aims to increase diversity in the aforementioned ecosystem, focusing on the implementation of GA4GH specifications, the native support for CWL workflows, the facilitation of the reproducibility of computational experiments, and the adaptive scheduling of jobs on heterogeneous clusters. The platform exploits containerization, experiment packaging, and workflow management technologies to ease reproducibility, while it leverages machine learning technologies to automatically identify the type of node that is more suitable to undertake each submitted computational task. 

It is worth mentioning that a deployment of SCHeMa powers the on-demand computations performed on the Cloud infrastructure of the ELIXIR-GR community\footnote{ELIXIR-GR Cloud Infrastructure (in beta): \url{https://egci-beta.imsi.athenarc.gr/}}, consisting of 45 physical nodes with 2600 CPU cores, 24 TBs RAM memory and 1 PB of storage, in total.

\section{System Overview}
\label{sec:overview}

\subsection{Design Objectives}
\label{sec:challenges}

SCHeMa, our open-source platform, has been designed and implemented with the aim to assist the work of scientists in the era of data-driven and reproducible science. In this context, our design had two main objectives: (a)~to make the reproduction of any computational experiment performed in the platform as easy as possible, and (b)~to allocate the resources of the underlying heterogeneous cluster as wisely as possible. 

Regarding the first objective, as was mentioned in Section~\ref{sec:intro}, a set of technologies should be combined together to achieve the desired goal. In addition, we had to identify the most appropriate technologies to be used from a multitude from available options. Our selection was made taking into consideration the maturity of the technologies to be used, their compatibility to each other, and the level of their dissemination in the scientific community. Based on these criteria, we selected to adopt CWL\footnote{CWL website: \url{https://www.commonwl.org/}} to describe software interfaces and workflows. We selected RO-crate~\cite{ro-crate} to create packages that represent computational experiments by storing the CWL description and the particular configuration of the software used along with the input and output datasets involved. Finally, we used containerization (Docker in particular) as a technology to enable the easy software execution, without requiring technical knowledge about building the involved software packages. 

Regarding the second objective, we approached the problem of selecting the most appropriate type of machine in the cluster as a classification problem where the input features are metadata relevant to the job to be executed (i.e., characteristics of the inputs used), while each class represents a particular type of machine (e.g., regular-memory machine, large-memory machine, slow-disk machine, etc). In particular, we implemented a profiler that, after a user request, is able to analyse the execution behavior of a software of interest by monitoring its execution on a wide range of different inputs and configurations. The profiler trains and evaluates the accuracy of various classification approaches in different hyperparameter configurations and selects the best performing one as the prediction model to be used for any execution of the particular software in the future.

\subsection{Architecture}
\label{sec:arch}

\begin{figure}[t]
  \centering
  \includegraphics[width=0.98\linewidth]{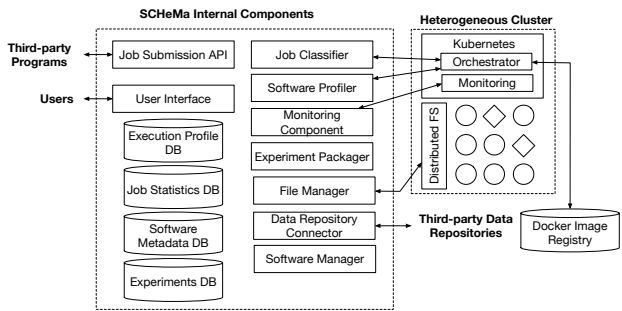}
  \caption{The architecture of SCHeMa.}
  \label{fig:arch}
\end{figure}

SCHeMa implements a wide range of functionalities to assist scientists in the data-driven and reproducible science era. Most notable are (a)~the option to upload custom-made scientific containers or container-based workflows, (b)~a wizard and an API that facilitate the execution of individual containers or workflows, (c)~a machine-learning-based classifier that (after a required training phase) identifies the type of cluster node which is more appropriate to undertake a particular computational task, (d)~a monitor that informs the users about the consumption of computational resources, (e)~a wizard to transform executed analyses into RO-crate-based ``experiment packages'', and (f)~a wizard to facilitate interconnection with open data repository services.
Figure~\ref{fig:arch} summarises SCHeMa's architecture, which supports these (and some extra) functionalities. In the following sections we discuss SCHeMa's external dependencies, as well as the implementation details of its internal software components.

\subsubsection{External dependencies}
\label{sec:dependencies}
SCHeMa's function depends on the existence of a couple of external installations, the most important of which are the following:

\begin{itemize}
    \item A \emph{Kubernetes\footnote{Kubernetes website: \url{https://kubernetes.io/}} installation} should be deployed on the computational cluster to be used. Kubernetes undertakes the low-level orchestration and monitoring of the computational jobs and interacts with the \emph{Job Classifier} component that provides the feed of requested jobs along with recommendations about the most suitable types of node for each job. 
    \item A \emph{distributed file system} should be installed on the hard disks of the machines of the cluster. This file system is used to store input/output data required/produced by the computational tasks. Currently the implementation supports NFS volumes.
    %, however, in the future other options will also be available (e.g., Ceph-based storage).  
    \item A private \emph{Docker image registry} supporting TLS security and user authentication is required. SCHeMa uses this registry to upload container images. This ensures that user-uploaded images remain isolated from the outside world (especially those meant to be private).
\end{itemize}

\subsubsection{User Interface}
\label{sec:webui}
A Web-based user interface has been developed using the Yii2 PHP framework\footnote{Yii2 website: \url{https://www.yiiframework.com/}}. It comprises various wizards that offer execution, reproducibility, and monitoring functionalities for computational experiments (see also Section~\ref{sec:ui}). Of course, the function of all these wizards heavily relies on the functionalities provided by the rest components, on which we elaborate in the next sections.

\subsubsection{Software Manager}
\label{sec:manager}

This component implements functionalities to upload (or update) container images and workflows. First of all, any involved container should be loaded in the \emph{Docker Image Registry}. In addition, in both cases (i.e., individual container or workflow), a CWL description is required and the corresponding metadata, which describe the required inputs and dependencies of the involved software packages (the latter only for workflows), are loaded in the \emph{Software Metadata DB}. These data are used by various components, e.g., they are used by the \emph{User Interface} wizard to automatically display a form containing one input field for each input parameter of the software (see also the example of Section~\ref{sec:webui}). 

\subsubsection{Job Submission API}
\label{sec:api}

Apart from manually submitting computational jobs through the UI wizards, the users are able to also submit batches of jobs programmatically using an implemented API. This API is based on GA4GH's Task Execution Schemas (TES)\footnote{TES specification: \url{https://github.com/ga4gh/task-execution-schemas}} and Workflow Execution Service (WES)\footnote{WES specification: \url{https://github.com/ga4gh/workflow-execution-service-schemas}} API specifications. The API supports batch execution and monitoring of computational tasks and can be used by user-implemented scripts and programs.

\subsubsection{Software Profiler}
\label{sec:profiler}

This component leverages machine learning to produce (after user request) ``execution profiles'' for software that has been uploaded on SCHeMa. For each software, this profiler builds a classification model that attempts to map candidate jobs of this software to one class of nodes, which corresponds to the type of node that is appropriate to undertake the computations of the job. 
As an indicative example, a cluster could have two types of nodes, one with regular-sized main memory and another one with large memory (ideally, to be used by memory-intensive tasks). In this case, given a particular software of interest, the objective of \emph{Software Profiler} would be to train a (binary) classification model to assign each job of this software to a regular-memory node or to a large-memory one. 

The profiling process goes as follows: first, the uploader of the software provides a set of alternative values/files for its input parameters (a relevant \emph{User Interface} wizard exists to collect this information). Then, the system runs the software for all input combinations, collects the resources consumption of each run, and creates a dataset of samples, where each sample is the combination of provided inputs and the recorded resources consumption. Based on this dataset, the system trains and optimizes different classifier models (implemented by scikit-learn\footnote{https://scikit-learn.org/stable/}) using the grid search approach. The model presenting the best accuracy is selected and stored in the \emph{Execution Profile DB}. All the stored models are exploited by the \emph{Job Classifier} component, when the execution of the corresponding software is requested.

\subsubsection{Job Classifier}
\label{sec:classifier}

This component receives as input job submissions from the corresponding wizard of the \emph{User Interface} and the \emph{Job Submission API}. 
As a first step, it searches if there are any trained (by the \emph{Software Profiler}) models for the involved software packages. If yes, it exploits this model to create a suggestion for the most appropriate node type to undertake the job and then propagates it to the Kubernetes scheduler through the Kubernetes API. The scheduler takes into consideration this suggestion and schedules the corresponding job ensuring that the determined node type will be used. If there is no trained model, then the job request is propagated to the Kubernetes shceduler without any indication for the node type to be used. 

\subsubsection{Data Repository Connector.}
\label{sec:data-connector}

This component implements an interconnection with various open data repositories. Currently, two repository services are supported: Zenodo\footnote{Zenodo open repository: \url{https://zenodo.org/}} and HELIX\footnote{Hellenic Data Service (HELIX) repository: \url{https://hellenicdataservice.gr}}. \emph{Data Repository Connector} takes advantage of the APIs provided by the repository services to download/upload datasets from/to them. A relevant \emph{User Interface} wizard exploits this functionality to enable users to download existing datasets from one of the supported repositories and then use them for their analyses or to directly upload the output of a particular analysis on a selected data repository. 

\subsubsection{Experiment Packager.}

This component is responsible to create ``experiment packages'' (according to the RO-crate~\cite{ro-crate} specification) from previously executed computational jobs, after user request. A relevant wizard is implemented in the \emph{User Interface} and, based on it, the users can easily create packages that incorporate several metadata for the selected experiment, such as the software used, its configuration, the input and output dataset, or even the DOI of a relevant publication. To collect the required information, the \emph{Experiment Packager} component communicates with the \emph{Job Statistics} and the \emph{Software Metadata} databases. The resulting packages are stored inside an \emph{Experiments DB} but it is also possible for the users to download the packages to their local computer storage. The easy creation of RO-crate packages is one of the main SCHeMa functionalities that facilitate computational experiment reproducibility. 

\subsubsection{Monitoring Component}
\label{sec:monitor}

This component aggregates data coming from the low-level logging and monitoring mechanisms of Kubernetes to create insightful reports about the jobs being executed in the cluster. All jobs, both those submitted through the UI and those submitted through the API are being considered. It also provides statistics about the load of the cluster. The component constantly communicates with the \emph{Job Statistics DB} to update its recorded information or to use it for the production of the aggregated statistics. Finally, it propagates data to the job execution wizard so the user can monitor each job's output and status. 

\subsection{Quick Tour of the Interface}
\label{sec:ui}

For the interest of space, in this section we indicatively describe only the main functionalities around the execution of scientific containers (however, the same functionalities for workflows are very similar). It should be highlighted that it is practically impossible for all the offered functionalities to be described.  

\begin{figure}[t]
  \centering
  \includegraphics[width=0.98\linewidth]{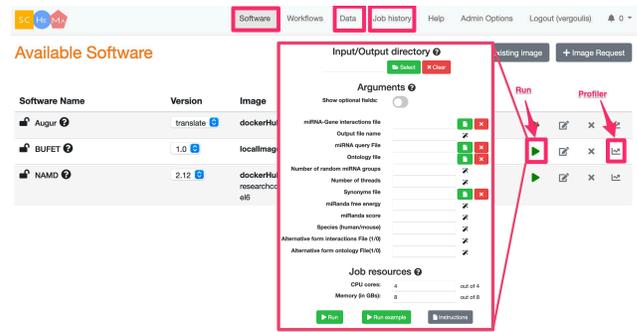}
  \caption{A screenshot of SCHeMa's interface.}
  \label{fig:screen}
\end{figure}

A list of all containers in the \emph{Docker Image Registry}, for which the connected user has access, can be displayed after clicking on the ``Software'' top menu item of SCHeMa. In the top of Figure~\ref{fig:screen}, a screenshot of SCHeMa's interface after this action took place is illustrated. By hitting the arrow-shaped button of any entry, the user is redirected to the job submission wizard of the corresponding software\footnote{Hitting the diagram-shaped button the user can start the machine learning profiling of the same software.}. This wizard consists of a form that contains one input field for each input parameter of the software; the form is automatically generated based on the CWL description of the selected software (which is stored in the \emph{Software Metadata DB}). After providing input values for all required fields and hitting ``Run'' the execution starts and the progress can be monitored through the UI. Closing the browser tab is possible without interrupting the execution; the user may revisit the execution page for this job by selecting the corresponding entry in the ``Job History'' page (accessible again through a top menu item). In the same page, the user can also select to rerun a previously completed execution or to create an RO-crate object based on it. Finally, any output files, which are stored in the distributed storage space, can be found in the ``Data'' page.

\section{Demonstration}
\label{sec:demo}

During the conference, we will explain the concept of reproducible and data-driven science and its requirements to the audience and we will demonstrate SCHeMa’s relevant functionalities elaborating on how they help in this context. For this demonstration, we will exploit SCHeMa's deployment for the ELIXIR-GR Cloud Infrastructure, which is based on a relatively large computational cluster (see Section~\ref{sec:intro}). We will examine SCHeMa's capabilities in real-time by following any audience-defined scenario, however we will also demonstrate some interesting scenarios we have identified. 

The main scenario, is based on executing a pre-loaded scientific container using the corresponding wizard. We will run the container twice: once without exploiting its pre-trained execution profile and once leveraging it and we will prompt the users to observe any differences (e.g., without using the profile a node with unnecessary large resources may be selected, thus its ``precious'' CPU time may be wasted instead of being used for a more demanding job). After the execution, we will also guide the audience through the process of uploading the output files on an open data repository and packaging the whole experiment into an RO-crate object. 

\section{Conclusion}
\label{sec:concl}

We introduced and demonstrated SCHeMa, an open-source platform that aims to assist the work of scientists in the data-driven science era through facilitating the execution, reproducibility, and monitoring of computational experiments on heterogeneous computational clusters. To this end, it leverages various technologies like containerization, experiment packaging, workflow management, and machine learning.

\begin{acks}
This work was partially funded by the ``ELIXIR-GR: Managing and Analysing Life Sciences Data (MIS: 5002780)'' project (co-financed by Greece and the European Union - European Regional Development Fund).
\end{acks}

\bibliographystyle{ACM-Reference-Format}
\bibliography{main}


\begin{thebibliography}{9}

%%% the Web address.

\ifx \showCODEN    \undefined \def \showCODEN     #1{\unskip}     \fi
\ifx \showDOI      \undefined \def \showDOI       #1{#1}\fi
\ifx \showISBNx    \undefined \def \showISBNx     #1{\unskip}     \fi
\ifx \showISBNxiii \undefined \def \showISBNxiii  #1{\unskip}     \fi
\ifx \showISSN     \undefined \def \showISSN      #1{\unskip}     \fi
\ifx \showLCCN     \undefined \def \showLCCN      #1{\unskip}     \fi
\ifx \shownote     \undefined \def \shownote      #1{#1}          \fi
\ifx \showarticletitle \undefined \def \showarticletitle #1{#1}   \fi
\ifx \showURL      \undefined \def \showURL       {\relax}        \fi
% invisible to TeX
\providecommand\bibfield[2]{#2}
\providecommand\bibinfo[2]{#2}
\providecommand\natexlab[1]{#1}
\providecommand\showeprint[2][]{arXiv:#2}

\bibitem[\protect\citeauthoryear{Afgan, Baker, Batut, van~den Beek, Bouvier,
  Cech, Chilton, Clements, Coraor, Gr{\"{u}}ning, Guerler, Hillman{-}Jackson,
  Hiltemann, Jalili, Rasche, Soranzo, Goecks, Taylor, Nekrutenko, and
  Blankenberg}{Afgan et~al\mbox{.}}{2018}]%
        {galaxy}
\bibfield{author}{\bibinfo{person}{Enis Afgan}, \bibinfo{person}{Dannon Baker},
  \bibinfo{person}{B{\'{e}}r{\'{e}}nice Batut}, \bibinfo{person}{Marius van~den
  Beek}, \bibinfo{person}{Dave Bouvier}, \bibinfo{person}{Martin Cech},
  \bibinfo{person}{John Chilton}, \bibinfo{person}{Dave Clements},
  \bibinfo{person}{Nate Coraor}, \bibinfo{person}{Bj{\"{o}}rn~A.
  Gr{\"{u}}ning}, \bibinfo{person}{Aysam Guerler}, \bibinfo{person}{Jennifer
  Hillman{-}Jackson}, \bibinfo{person}{Saskia~D. Hiltemann},
  \bibinfo{person}{Vahid Jalili}, \bibinfo{person}{Helena Rasche},
  \bibinfo{person}{Nicola Soranzo}, \bibinfo{person}{Jeremy Goecks},
  \bibinfo{person}{James Taylor}, \bibinfo{person}{Anton Nekrutenko}, {and}
  \bibinfo{person}{Daniel~J. Blankenberg}.} \bibinfo{year}{2018}\natexlab{}.
\newblock \showarticletitle{The Galaxy platform for accessible, reproducible
  and collaborative biomedical analyses: 2018 update}.
\newblock \bibinfo{journal}{\emph{Nucleic Acids Res.}} \bibinfo{volume}{46},
  \bibinfo{number}{Webserver-Issue} (\bibinfo{year}{2018}),
  \bibinfo{pages}{W537--W544}.
\newblock
\urldef\tempurl%
\url{https://doi.org/10.1093/nar/gky379}
\showDOI{\tempurl}


\bibitem[\protect\citeauthoryear{Baker}{Baker}{2016}]%
        {baker2016}
\bibfield{author}{\bibinfo{person}{Monya Baker}.}
  \bibinfo{year}{2016}\natexlab{}.
\newblock \showarticletitle{1,500 scientists lift the lid on reproducibility}.
\newblock \bibinfo{journal}{\emph{Nature}}  \bibinfo{volume}{533}
  (\bibinfo{year}{2016}), \bibinfo{pages}{452--454}.
\newblock
\urldef\tempurl%
\url{https://doi.org/10.1038/533452a}
\showDOI{\tempurl}


\bibitem[\protect\citeauthoryear{Boettiger}{Boettiger}{2014}]%
        {Boettiger14}
\bibfield{author}{\bibinfo{person}{Carl Boettiger}.}
  \bibinfo{year}{2014}\natexlab{}.
\newblock \showarticletitle{An introduction to Docker for reproducible
  research, with examples from the {R} environment}.
\newblock \bibinfo{journal}{\emph{CoRR}}  \bibinfo{volume}{abs/1410.0846}
  (\bibinfo{year}{2014}).
\newblock
\showeprint[arxiv]{1410.0846}
\urldef\tempurl%
\url{http://arxiv.org/abs/1410.0846}
\showURL{%
\tempurl}


\bibitem[\protect\citeauthoryear{Carrag{\'a}in, Goble, Sefton, and
  Soiland-Reyes}{Carrag{\'a}in et~al\mbox{.}}{2019}]%
        {ro-crate}
\bibfield{author}{\bibinfo{person}{Eoghan~{\'O} Carrag{\'a}in},
  \bibinfo{person}{Carole Goble}, \bibinfo{person}{Peter Sefton}, {and}
  \bibinfo{person}{Stian Soiland-Reyes}.} \bibinfo{year}{2019}\natexlab{}.
\newblock \showarticletitle{A lightweight approach to research object data
  packaging}. In \bibinfo{booktitle}{\emph{Bioinformatics Open Source
  Conference (BOSC) 2019}}.
\newblock


\bibitem[\protect\citeauthoryear{Chamberlain and Schommer}{Chamberlain and
  Schommer}{2014}]%
        {chamberlain2014}
\bibfield{author}{\bibinfo{person}{Ryan Chamberlain} {and}
  \bibinfo{person}{Jennifer Schommer}.} \bibinfo{year}{2014}\natexlab{}.
\newblock \showarticletitle{Using Docker to support reproducible research}.
\newblock \bibinfo{journal}{\emph{DOI: https://doi. org/10.6084/m9. figshare}}
  \bibinfo{volume}{1101910} (\bibinfo{year}{2014}), \bibinfo{pages}{44}.
\newblock


\bibitem[\protect\citeauthoryear{da~Veiga~Leprevost, Gr{\"u}ning,
  Alves~Aflitos, R{\"o}st, Uszkoreit, Barsnes, Vaudel, Moreno, Gatto, Weber,
  et~al\mbox{.}}{da~Veiga~Leprevost et~al\mbox{.}}{2017}]%
        {da2017biocontainers}
\bibfield{author}{\bibinfo{person}{Felipe da Veiga~Leprevost},
  \bibinfo{person}{Bj{\"o}rn~A Gr{\"u}ning}, \bibinfo{person}{Saulo
  Alves~Aflitos}, \bibinfo{person}{Hannes~L R{\"o}st}, \bibinfo{person}{Julian
  Uszkoreit}, \bibinfo{person}{Harald Barsnes}, \bibinfo{person}{Marc Vaudel},
  \bibinfo{person}{Pablo Moreno}, \bibinfo{person}{Laurent Gatto},
  \bibinfo{person}{Jonas Weber}, {et~al\mbox{.}}}
  \bibinfo{year}{2017}\natexlab{}.
\newblock \showarticletitle{BioContainers: an open-source and community-driven
  framework for software standardization}.
\newblock \bibinfo{journal}{\emph{Bioinformatics}} \bibinfo{volume}{33},
  \bibinfo{number}{16} (\bibinfo{year}{2017}), \bibinfo{pages}{2580--2582}.
\newblock


\bibitem[\protect\citeauthoryear{Fanelli}{Fanelli}{2018}]%
        {Fanelli2628}
\bibfield{author}{\bibinfo{person}{Daniele Fanelli}.}
  \bibinfo{year}{2018}\natexlab{}.
\newblock \showarticletitle{Opinion: Is science really facing a reproducibility
  crisis, and do we need it to?}
\newblock \bibinfo{journal}{\emph{Proceedings of the National Academy of
  Sciences (PNAS)}} \bibinfo{volume}{115}, \bibinfo{number}{11}
  (\bibinfo{year}{2018}), \bibinfo{pages}{2628--2631}.
\newblock
\urldef\tempurl%
\url{https://doi.org/10.1073/pnas.1708272114}
\showDOI{\tempurl}


\bibitem[\protect\citeauthoryear{Jimenez, Maltzahn, Moody, Mohror, Lofstead,
  Arpaci-Dusseau, and Arpaci-Dusseau}{Jimenez et~al\mbox{.}}{2015}]%
        {jimenez2015}
\bibfield{author}{\bibinfo{person}{Ivo Jimenez}, \bibinfo{person}{Carlos
  Maltzahn}, \bibinfo{person}{Adam Moody}, \bibinfo{person}{Kathryn Mohror},
  \bibinfo{person}{Jay Lofstead}, \bibinfo{person}{Remzi Arpaci-Dusseau}, {and}
  \bibinfo{person}{Andrea Arpaci-Dusseau}.} \bibinfo{year}{2015}\natexlab{}.
\newblock \showarticletitle{The role of container technology in reproducible
  computer systems research}. In \bibinfo{booktitle}{\emph{2015 IEEE
  International Conference on Cloud Engineering}}. IEEE,
  \bibinfo{pages}{379--385}.
\newblock


\bibitem[\protect\citeauthoryear{Wolstencroft, Owen, Krebs, Nguyen, Stanford,
  Golebiewski, Weidemann, Bittkowski, An, Shockley, Snoep, Müller, and
  Goble}{Wolstencroft et~al\mbox{.}}{2015}]%
        {SEEK2015}
\bibfield{author}{\bibinfo{person}{Katherine Wolstencroft},
  \bibinfo{person}{Stuart Owen}, \bibinfo{person}{Olga Krebs},
  \bibinfo{person}{Quyen Nguyen}, \bibinfo{person}{Natalie~J. Stanford},
  \bibinfo{person}{Martin Golebiewski}, \bibinfo{person}{Andreas Weidemann},
  \bibinfo{person}{Meik Bittkowski}, \bibinfo{person}{Lihua An},
  \bibinfo{person}{David Shockley}, \bibinfo{person}{Jacky~L. Snoep},
  \bibinfo{person}{Wolfgang Müller}, {and} \bibinfo{person}{Carole Goble}.}
  \bibinfo{year}{2015}\natexlab{}.
\newblock \showarticletitle{SEEK: a systems biology data and model management
  platform}.
\newblock \bibinfo{journal}{\emph{BMC Systems Biology}} \bibinfo{volume}{9},
  \bibinfo{number}{1} (\bibinfo{date}{11 Jul} \bibinfo{year}{2015}),
  \bibinfo{pages}{33}.
\newblock
\showISSN{1752-0509}
\urldef\tempurl%
\url{https://doi.org/10.1186/s12918-015-0174-y}
\showDOI{\tempurl}


\end{thebibliography}

\end{document}